\documentclass[12pt,a4paper]{article}

\usepackage{latexsym}
\usepackage{amsmath}
\usepackage{amsfonts}
\usepackage{amssymb}
\usepackage{amsthm}
\usepackage{amscd}
\usepackage{mathrsfs}

\def\bC{{\mathbb C}}                    %
\def\be{\begin{equation}}              %
\def\ee{\end{equation}}                %
\def\no{\noindent}                      %
\def\om{\omega}

\def\bR{{\mathbb R}}                    %
\def\bC{{\mathbb C}}                    %
\def\bZ{{\mathbb Z}}                    %
 
\def\A{{\cal A}}
\def\F{{\cal F}}

\def\w{\wedge}

\def\sld{sl(2\vert 1,\bC)}

\def\uosp{uosp(2\vert 1)}
\def\st{\stackrel{\otimes}{,}}
\def\Tr{{\rm Tr}}
\def\ST{{\rm STr}}
\def\ss{\subset}

\def\de{\delta}
\def\sus{S^{2\vert 2}}

\def\dd{\ddagger}

\def\jp{\frac{1}{2}}
\def\js{\frac{1}{4}}

\def\1{{\mbox{\boldmath $1$}}}  

\def\be{\begin{equation}}
\def\ee{\end{equation}}
\def\bea{\begin{eqnarray}}
\def\eea{\end{eqnarray}}
\def\apo{C^{pol}(S^2)}
\def\sapo{C^{pol}(S^{2\vert 2})}

\def\M{{\cal M}}
\def\U{{\cal U}}
\def\V{{\cal V}}
\def\S{{\cal S}}

\def\tro{d\mu_{S^2}}
\def\std{d\mu_{S^{2\vert 2}}}

\def\*{\ddagger}

\def\Ad{{\rm Ad}}
\def\vv{\vec{\times}}
\def\ri{{\mathrm{i}}}    
\def\vph{\vec{\Phi}}
\def\vps{\vec{\Psi}}

\def\e{\varepsilon}

\def\tt{{\bf\times}}

\def\bph{{\bf \Phi}}
\def\bps{{\bf \Psi}}
\def\bu{{\bf u}}
\begin{document}

\newtheorem{theorem}{Theorem}[section]
\begin{titlepage}
\vspace*{0.5cm}
\begin{center}
{\Large \bf On Poisson  geometry and supersymmetric sigma models }

\end{center}

\vspace{0.2cm}

\begin{center}
  C. Klim\v c\'\i k  \\

\bigskip

 Institut de math\'ematiques de Luminy,
 \\ 163, Avenue de Luminy, \\ 13288 Marseille, France\\
 e-mail: ctirad.klimcik@univmed.fr

\bigskip

\end{center}

\vspace{0.2cm}

\begin{abstract} 
By using the Poisson geometry, we develop   a manifestly  invariant and calculation-friendly formalism for handling  $UOSp(2\vert 1)$-supersymmetric field theories. In particular, the super-Langrangians are written solely in terms of superfields, Poisson brackets and the  moment map generating the $UOSp(2\vert 1)$ action. As an application of this formalism, 
we construct the Kalb-Ramond term  for supersymmetric sigma models on the supersphere.   \end{abstract}

\end{titlepage}
\newpage

\section{Introduction}
Consider a  smooth map $\phi:\Sigma\to T$ where $\Sigma$ and $T$ are Riemannian manifolds. Such a map is called harmonic if it is a solution of field equations
of the so called nonlinear sigma model associated to $\Sigma$ and $T$. The Lagrangian of this model is given by the squared double  norm (with respect to the metrics $g$ on $\Sigma$ and $G$ on $T$)  of the derivation of $\phi$ and the action is obtained by integration of the Lagrangian with respect to the measure $d\mu_g$ on $\Sigma$ induced by the metric $g$:
\be \S_G
= \int d\mu_g \vert\vert d\phi\vert\vert_{g,G}^2.\label{GB}\ee
Obviously, the nonlinear sigma model is symmetric with respect to the
group of isometries of the source manifold $\Sigma$. For example, if $\Sigma$ is the two sphere $S^2$ the model has the rotational $SO(3)$ symmetry.

If $\Sigma$ is two-dimensional, there exists  a generalisation of the nonlinear sigma model considered mainly in string theory which is induced by a presence of an additional
geometrical structure on the target $T$. This  structure is called the Kalb-Ramond field \cite{KR} and it is nothing but a two-form field $B$ on $T$.
The pull-back $\phi^*B$ integrated over $\Sigma$ is then added to the original sigma model action in order to take into account the presence of
$B$:
\be \S_{GB}
= \int d\mu_g \vert\vert d\phi\vert\vert_{g,G}^2+\int \phi^*B.\label{Kal}\ee
Note that the Kalb-Ramond term $\int \phi^*B$ is not only invariant with respect to the isometries of $\Sigma$ but it is invariant  even with respect
to all diffeomorphisms of $\Sigma$.

If $E$ is a Riemannian supermanifold  with the bosonic body being the flat Euclidean $2$-plane and odd coordinates being  $\xi,\bar\xi$ then there exists  a supersymmetric generalisation
of the nonlinear sigma model  which includes the Kalb-Ramond term \cite{HS,CZ}.  Its 	action in the form of the Berezin integral reads
\be S_E=\int d\bar z  dz\bar d\xi d\xi(G_{IJ}(Y^K)+ \ri B_{IJ}(Y^K)) DY^I  \bar DY^J,\label{euc}\ee
where $Y^I$ are the superfields (i.e. even functions on the Euclidean 
superplane) corresponding to  the coordinates  on the target space and $G_{IJ}$ and $B_{IJ}$ are respectively the components of the target space metric and the
Kalb-Ramond field in those coordinates. Moreover, the supersymmetric covariant derivatives are defined as
\be \bar D:=\partial_{\bar\xi} +\bar\xi\partial_{\bar z},\quad D:=\partial_\xi+\xi\partial_z.\ee
Now the Kalb-Ramond part of the  supersymmetric  action (\ref{euc}) is a less geometric object as in the bosonic case since it  can  no longer be written in terms of  a pull-back   $Y^*B$ by a sigma model superfield $Y$. Indeed,  though  $B$ is the two-form, the volume form on the source supermanifold $E$ is not a two-form  
due to the presence of odd differentials. 
As the result, the  Kalb-Ramond term is not supersymmetric with respect to all superdiffeomorphisms of $E$ but just with respect to
the superisometries of $E$. In fact, the supersymmetric Kalb-Ramond term is  {\it determined} by the criterion of superinvariance with respect to the Euclidean superisometries and by the criterion that, when the   superfields $Y^I$  do not depend on the odd coordinates $\bar \xi$,$\xi$,   the supersymmetric action (\ref{euc})
must reduce to the bosonic action (\ref{GB}). 

The arguments of the previous paragraph   show that in the supersymmetric case the Kalb-Ramond term
must be determined case by case following what is the geometry  of, possibly curved, world-sheet. More precisely, the Kalb-Ramond term is to be  determined by two conditions: it must be invariant with respect to the superisometries of the worldsheet and it must reduce to the geometric term $\int\phi^*B$ in the bosonic limit. In this paper, we shall consider
sigma models on 
the so called supersphere \cite{GKP} which is the simplest  supersymmetrization  of the standard sphere with the supergroup of superisometries   being the  unitary orthosymplectic supergroup $UOSp(2\vert 1)$.
In fact, to construct   the action of the $UOSp(2\vert 1)$ supersymmetric nonlinear sigma model on the curved super-worldsheet $\sus$
is far from being just a straightforward generalisation of the flat super-Euclidean situation. This  problem  was posed already  in \cite{GKP} but it is fully solved only in the present article since the old
  work  \cite{GKP}  and the subsequent work \cite{SW}  constructed    the supersymmetric sigma model on $\sus$     without the Kalb-Ramond term.

Speaking more generally, string theoretical motivations caused recently  a growing interest  in formulations of rigidly supersymmetric  field theories  on curved space-times \cite{AJKL,BNY,BK,GKL,FS, H,KMMR,LM,P,R,ST}.
 Obviously, the principal examples of such space-times are homogeneous spaces
of    supergroups e.g.  $\sus$ itself   can be understood as the coset supermanifold $UOSp(2\vert 1)/U(1)$. Although it is quite straightforward to construct various differential supersymmetric invariants of the rigid supersymmetry supergroup it is fairly less trivial task to work out which invariants give rise to viable field theories.
 Indeed, a seemingly "nice" invariant action principle written in the superfield formalism  may be pathological when worked out in components. Typically, there may occur a violation of spin-statistics (a presence of quadratic  bosonic derivatives  in the fermionic kinetic term)  and also
 other unwanted phenomena  (like fourth order bosonic  derivatives   in the case the theory contains a gauge symmetry).    Clues  to select non-pathological  candidates vary from case to case and no universal algorithms are available.  
 
In the search of  a viable definition
of the supersymmetric sigma model on the super-worldsheet $\sus$ we take a profit of the existence of a natural Poisson structure on $\sus$ .   In fact,
 the Poisson insights in the search for consistent supersymmetric field theories were not so far much  explored in the literature  though  the original paper \cite{GKP} using
 this approach   is around already for some time.  We believe that it's a pity because many expressions become less heavy when written in the Poisson way. Indeed,  the principal result of this paper, which is  the construction of the $UOSp(2\vert 1)$ invariant  supersymmetric 
 sigma model action including the Kalb-Ramond term on the supersphere, is obtained  by relying on Poisson geometry   both as the source of inspiration and  
 as  a  technical tool. Without anticipating all details, this action reads
   \be S_{sGB}=-\ST\int \std (G_{IJ} +2\ri\M B_{IJ} ) \{\M^2,Y^I\}\{\M^2,Y^J\},\label{cor0}\ee
   where (the moment map) $\M$ is a fixed   $\uosp$-matrix valued function on $\sus$, $\{.,.\}$ stands for the Poisson bracket and
   $\std$ is an $\uosp$-invariant measure on $\sus$. We note the appearance in the action (\ref{cor0}) of the both  moment map $\M$, which
   encompasses all supersymmetric generators  and its square $\M^2$, which turns out to encompass  all supersymmetric covariant derivatives.
   It is precisely this circumstance that illustrates that from the structural point of view the  supersymmetric action (\ref{cor0}) is not quite a direct generalisation of the purely bosonic $SO(3)$ invariant sigma model the action of which reads
      \be S_{GB}=\frac{1}{2}\Tr\int \tro (G_{IJ} +\ri M B_{IJ} ) \{M,Y^I\}\{M,Y^J\}.\label{bcor}\ee
      Here the bosonic moment map $M$ generates the $SO(3)$ symmetry and its square does not appear in the story (in fact, unlike
      the  moment map $\M$, $M$  squares to the unit matrix!). Inspite of differences,  the supersymmetric action (\ref{cor0}) will be shown to reduce to the bosonic action (\ref{bcor}) when  the fermionic parts of the superfields $Y^I$ are set to zero.

In a short Section 2, we  review the concept of a  Hermitian supermatrix and then in  Section 3 we describe features of the Poisson geometries of the sphere $S^2$ and of  the supersphere $\sus$. Finally, in Section 4, 
we first construct  the ordinary $SO(3)$ invariant bosonic sigma model (\ref{bcor}) on the ordinary sphere $S^2$, then the $UOSp(2\vert 1)$-invariant  super sigma model (\ref{cor0})  on the supersphere $\sus$ and  we establish that in the absence of the fermions the
superaction (\ref{cor0}) does reduce to the bosonic action (\ref{bcor}). 
 
\section{Supermatrices}
Consider a complex  Grassmann algebra $G$ equipped with  a $\bC$-antilinear map call graded conjugation \cite{RS}, which associates to every $a\in G$ an element $\bar a\in G$ in such a way that
\be \overline{ab}=\bar a\bar b,  \quad \bar{\bar a}=(-1)^{p(a)}a,   \qquad a,b\in G.\ee
Here $p(a)$ means the Grassmann parity of $a$.
By  a supermatrix\footnote{In this paper we shall not consider "odd" supermatrices for which   the Grassmann parity
of an element $M_{ij}$ is opposite to   the sum $p(i)+p(j)$ of the index parities} we mean a square  matrix  $M$ with a distinguished parities of indices    for which
 the Grassmann parity
of an element $M_{ij}\in G$ is the same as   the sum $p(i)+p(j)$ of the index parities. Moreover,  the elements $M_{ij}$ of  a "Hermitian supermatrix"  satisfy  the relation
\be M_{ij}=\bar M_{ji},\quad i \geq j.\label{her} \ee
Note that in the purely bosonic  case  (\ref{her}) remains true for all indices $i,j$, however in the supercase the restriction
to the  inequality $i\geq j$ is essential.

\no The supertrace $\ST(M)$ of a supermatrix $M$   is defined as
\be \ST(M):=\sum_i(-1)^{p(i)}M_{ii}.\ee
There is a natural  supermeasure  $d\mu_H$ on the superhermitian matrices given by the formula
\be d\mu_H:= \Pi_idM_{ii}\Pi_{i<j}dM_{ij}d\bar M_{ij},\label{hm}\ee
where $dM_{ij}d\bar M_{ij}$ is the  Berezin measure if $p(i)+p(j)$ is odd and the standard Lebesgue measure on $\bC=\bR^2$ if $p(i)+p(j)$ is even.

The purely "bosonic" case corresponds to the situation where all index parities are even. All formulae presented in this section remains then true  just the terminology flips  e.g. from the supertrace  $\ST$ to
 the ordinary trace $\Tr$ etc.

\section{Sphere and supersphere}
\subsection{Sphere $S^2$}
We describe the ordinary two-dimensional sphere in the  way best suited for the later supersymmetric generalisation.  Thus
we define the sphere $S^2$ as the set of ordinary (purely bosonic)  Hermitian $2\times 2$ matrices $M$ such that
\be \Tr(M)=0, \quad \Tr(M^2)=2.\label{sph}\ee
Indeed, in terms of the matrix components $M_{ij}$     the first condition gives
\be M_{11}=-M_{22}\ee
and  the second one
\be M_{11}^2+\bar M_{12} M_{12}=1.\label{S}\ee
We shall view the Hermitian matrices $M$ verifying the conditions (\ref{sph}) as points on the sphere but the matrix elements $M_{ij}$
as particular {\it functions} on the sphere.  The algebra $\apo$ generated by $M_{ij}$  is then a (dense) subspace  of the space $C^\infty(S^2)$ of smooth complex
functions on $S^2$.

\no The sphere (e.g.  viewed as the surface of the unit ball in the three-dimensional Euclidean space)  can be naturally
 rotated by the group $SO(3)$. The infinitesimal generators  $V\in so(3)$ of this action turn out  to act on the point  $M$ of the sphere 
 as $\ri[V,M]$, where $V$ is viewed as a traceless Hermitian matrix.  (The Lie commutator on $so(3)$ is then $\ri$-multiple of the matrix commutator.)

\no A natural Poisson bracket on $\apo$  is  defined by the following formula
\be \{\Tr(UM),\Tr(VM)\}:=-\ri\Tr([U,V]M),\label{pb}\ee
where $U,V$ are  any  constant traceless Hermitian matrices and $[U,V]$  is the standard matrix commutator.
The defining brackets (\ref{pb}) can be rewritten equivalently as
\be \{\Tr(UM),M\}=\ri[U,M],\label{pb1}\ee
 It is easy to verify that (\ref{pb})  indeed determines a Poisson bracket, in particular, the Poisson Jacobi identity is the consequence of the matrix Jacobi idenity.  Moreover, by taking the trace of (\ref{pb1}) and of $ \{ \Tr(UM),M^2\}=\ri[U,M^2]$, we derive
\be \{M,\Tr(M)\}=\{M,\Tr(M^2)\}=0,\label{inv}\ee which is obviously needed for consistence with the definition (\ref{sph}) of the sphere.

\no  Looking at (\ref{pb1}), we immediately see
 that the $so(3)$-action is Hamiltonian with respect to the Poisson structure $\{.,.\}$. The corresponding moment map   is clearly $M$ and the Hamiltonian  corresponding to the $so(3)$ generator $U$ is $\Tr(UM)$. It follows   that the Poisson structure (\ref{pb}) is $so(3)$ invariant:
 \be \{\Tr(UM),\{f,g\}\}= \{\{\Tr(UM),f\}g\}+\{f,\{\Tr(UM),g\}\}, \quad \forall f,g\in\apo.\ee
 A natural round measure on the sphere $S^2$ can be defined with the help of the measure $d\mu_H$ on Hermitian matrices weighted
by  delta functions of the  constraints which define the sphere:
\be \tro :=  d\mu_H\delta(\Tr M)\delta(\jp\Tr M^2-1)= dM_{11}dM_{12}d\bar M_{12}\delta(M_{11}^2+\bar M_{12} M_{12}-1).\ee
We now wish to check, that  this measure $\tro$  is indeed rotational invariant. For that it is sufficient to check the invariance of $d\mu_H$
since the invariance of the arguments of the delta functions follows from (\ref{inv}).  The infinitesimal change of coordinates induced by the rotation $V$ is obviously
\be \delta M=i\e [V,M]\equiv i \e  \Ad_VM\ee
where $\e$ is a small parameter.
The induced Jacobian is then 
\be {\rm det}(1+i \e  {\rm Ad}_V)=1+i \e \Tr(\Ad_V)=1,\ee
which means that the measure is indeed invariant.

\no The immediate consequence of the invariance of the measure $\tro$ is the formula
\be \int \tro \{M,f\}=0, \quad \forall f\in \apo,\ee
since $\{M,f\}$ is the (matrix valued) variation of the function $f$ under (all possible) infinitesimal rotations.

 \subsection{Supersphere $S^{2\vert 2}$}
 A $3\times 3$ Hermitian supermatrix $\V$  with two even indices $1,2$ and one odd index $3$ is called orthosymplectic, if it satisfies
  \be \V_{33}=0,\quad \V_{23}=\bar \V_{13}.\label{ort}\ee
  An $\ri$-multiple of the standard  commutator of two orthosymplectic supermatrices is again orthosymplectic and the corresponding
(unitary orthosymplectic)  Lie superalgebra is referred to as $\uosp$.

\no We now define the supersphere (or rather the algebra $\sapo$ of  polynomial functions on the supersphere) in a more invariant way 
than in  \cite{GKP}, namely, we view it as the algebra    generated by matrix elements  of a Hermitian orthosymplectic supermatrix $\M$  submitted to further
constraints
\be \ST(\M)=0,\quad \ST(\M^2)=2.\label{ssp}\ee
Equivalently, solving the   linear constraints give five independent generators   which must verify the remaining quadratic constraint \be \M_{11}^2+ \M_{12} \bar\M_{12}+2 \M_{13}\bar\M_{13}=1.\label{SSP}\ee
Note, that if the odd generators $\M_{13},\bar \M_{13}$ vanish then (\ref{SSP}) reduces to the defining relation (\ref{S}) of the ordinary 
sphere.

The supersphere   can be "superrotated" by the unitary orthosymplectic group $UOSp(2\vert 1)$ the Lie superalgebra of which is 
$\uosp$. Infinitesimal action of $\V\in\uosp$ is just given by $\ri[\V,\M]$. 
This    action is Hamiltonian  (with the Hamiltonian equal to $\ST(\V\M)$  and 
  the moment map equal to $\M\in \uosp$) if we define
an $\uosp$ invariant Poisson structure on $\sapo$ by the bracket
\be \{\ST(\U\M),\ST(\V\M)\}:=-\ri \ST([\U,\V]\M), \quad \U,\V\in\uosp.\label{spb1}\ee
Indeed, (\ref{spb1}) clearly implies
\be \{ \ST(\V\M),\M\}=i[\V,\M].\label{spb}\ee
It  can  be  also easily checked that it holds
\be \{\M,\ST(\M)\}=\{\M,\ST(\M^2)\}=0\ee
as the consistency requires.

\no A natural  $\uosp$ invariant measure on the supersphere $S^{2\vert 2}$ can be defined with the help of the measure (\ref{hm}) on Hermitian supermatrices weighted
by  delta functions of all  constraints which define the supersphere:
$$ \std := d\mu_H\delta(\ST \M)\delta(\jp\ST \M^2-1)\delta(\M_{33})\delta(\M_{23}-\bar\M_{13})\delta(\bar\M_{23}+ \M_{13})=$$
\be = d\M_{11}d\M_{12}d\bar \M_{12}d\M_{13}d\bar\M_{13}\delta(\M_{11}^2+ \M_{12}  \bar\M_{12}+2\M_{13}  \bar\M_{13}-1).\ee
Due to $\uosp$ invariance of the constraints,  in order to check the invariance of the measure $\std$, it is  sufficient to check the invariance of $d\mu_H$.  The infinitesimal change of coordinates induced by the $\uosp$ element   $\V$ is obviously
\be \delta \M=i\e [\V,\M]\equiv i \e  \Ad_{\V}\M\ee
where $\e$ is a small parameter.
The induced Berezinian is then 
\be {\rm sdet}(1+i \e  {\rm Ad}_{\V})=1+i \e \ST(\Ad_{\V})=1,\ee
which means that the measure is indeed $\uosp$ invariant.

\no The immediate consequence of the invariance of the measure $\std$ is the formula
\be \int \std \{\M,f\}=0, \quad \forall f\in \sapo,\ee
since $\{\M,f\}$ is the (matrix valued) variation of the function $f$ under (all possible) infinitesimal $\uosp$ transformations.

\section{Sigma models}

\subsection{The bosonic case}

 Denote by $y^I$, $I=1,...,n$ coordinates on the target Riemannian manifold $T$  and, slightly abusing the notation, also the pull-backs $\phi^*y^I$ by some smooth map $\phi:S^2\to T$. The bosonic sigma model action $\int d\mu_g \vert\vert d\phi\vert\vert_{g,G}^2$  with the standard
round metric $g$ on $S^2$ and a metric $G_{IJ}(y^K)$ on $T$  can be then rewritten in the following way (cf. Eq. (134) of \cite{GKP}):
\be S_G=\int \tro G_{IJ}(y^K)\{x_m,y^I\}\{x_m,y^J\}.\label{bosg}\ee
Here    $\{.,.\}$ is the   Poisson structure defined in (\ref{pb}) and
$x_m$ are the fixed "Cartesian" functions on the sphere defined by the standard embedding of $S^2$ into the Euclidean space $\bR^3$. In the 
notation  of Section 3.1 we have $M=x_m\sigma^m$ (where $\sigma^m$ are the standard Pauli matrices)  and   we can therefore rewrite (\ref{bosg}) in more invariant way as
\be S_{G}=\jp\Tr\int \tro G_{IJ}(y^K) \{M,y^I\}\{M,y^J\} .\label{bosg1}\ee
It may appear natural to add to (\ref{bosg1}) the Kalb-Ramond term in its most symmetric form $\int \phi^*B$, however, such expression does not lend
itself to the supersymmetric generalisation. We shall instead  rewrite the full sigma model action $S_{GB}=$$\int d\mu_g \vert\vert d\phi\vert\vert_{g,G}^2$ $+$ $\int \phi^*B$    as 
\be S_{GB}=\jp\Tr\int \tro (G_{IJ} +\ri MB_{IJ} ) \{M,y^I\}\{M,y^J\}\equiv \Tr\int \tro L_{GB},\label{Kal1}\ee
where $B_{IJ}$ are the components of the Kalb-Ramond form $B$ in the coordinates $y^I$. The representation (\ref{Kal1}) of the full
sigma model action was not obtained in \cite{GKP}, therefore we have to justify it.   For that we must first check
  the $so(3)$ invariance of the action (\ref{Kal1}) with respect to the infinitesimal rotations $\delta_V y^I=\{\Tr(VM),y^I\}$. Using (\ref{pb1}),
  we find successively
  \be \delta_VG_{IJ}(y^K)=\{\Tr(VM), G_{IJ}(y^K)\},\label{a}\ee
  \be \delta_V(B_{IJ}(y^K)M)=\{\Tr(VM), B_{IJ}(y^K)\}M=\{\Tr(VM), B_{IJ}(y^K)M\}-\ri[V, B_{IJ}(y^K)M],\label{b}\ee
  \be \delta_V\{M,y^I\}=\{M,\delta_Vy^I\}= \{\Tr(VM), \{M,y^I\}\}-i[V,\{M,y^I\}],\label{c}\ee
  \be \delta_V S_{GB}=\Tr\int\tro\{\Tr(VM),L_{GB}\}-\ri\Tr\int\tro[V, L_{GB}]=0.\label{d}\ee
  Indeed, the last equality follows from (21) and  from the  fact that  $\Tr VL_{GB}=\Tr L_{GB}V$.
  Now we are ready to verify that the Kalb-Ramond term $\int \phi^*B$ can be written as 
  \be \int \phi^*B=\jp\Tr\int \tro \ri MB_{IJ} (y^K)  \{M,y^I\}\{M,y^J\}.\label{pe}\ee
 Indeed the integral of the differential form $\phi^*(dy^I\wedge dy^J)$ over $S^2$ can be certainly  written as $\int \tro <K,dy^I\wedge dy^J>$ where $K$ is some bivector on $S^2$ (we write $y^I$ instead of $\phi^*y^I$). Because $S^2$ is two-dimensional manifold, every
 two bivectors $K$ and $\tilde K$ are related as $\tilde K=fK$, where $f$ is a function on the sphere. This means, in particular, that
   \be \int \phi^*B=\jp\Tr\int \tro f\ri MB_{IJ} (y^K)  \{M,y^I\}\{M,y^J\}.\label{pe2}\ee
 However, the $so(3)$ invariance of both   $\int \phi^*B$ and $\Tr\int \tro \ri MB_{IJ} (y^K)  \{M,y^I\}\{M,y^J\}$ means that $f$ must be a $so(3)$ invariant function hence a constant and it is easy to check that $f=1$.

\subsection{The supersymmetric case}
In analogy with the bosonic case (\ref{Kal1}), it looks plausible that the action of the supersymmetric sigma model should  
be of the type
\be S_{tent}=\ST\int \std (G_{IJ} +\ri\M B_{IJ} ) \{\M,Y^I\}\{\M,Y^J\}.\label{tent}\ee
Here $Y^I$ are the sigma model superfields viewed as elements of $\sapo$, all other symbols were introduced in Section 3.2, and we wrote $S_{tent}$ to indicate that this is just the tentative expression.
Indeed, quite remarkably,  this  caution turns out to be fully  justified since the tentative action (\ref{tent}) is pathological when worked out in components! (The problem is that upon the expansion in components the kinetic term for the fermions contains two bosonic derivatives.)  It is therefore necessary to look for  another expression  and the main result of this article states that such
a viable  $\uosp$-invariant action is in fact given by the following formula
\be S_{sGB}=-\ST\int \std (G_{IJ} +2\ri\M B_{IJ} ) \{\M^2,Y^I\}\{\M^2,Y^J\}.\label{cor}\ee
There are three things that we have to verify in order to show that the action (\ref{cor}) is indeed the correct one. First of all it is
$\uosp$-invariance,  then the correct bosonic limit and, thirdly, the absence  of a quadratic expression in the bosonic derivatives in the fermionic part of the action. 

\noindent 1. The $\uosp$-invariance  of (\ref{cor})  is verified in the exactly same way (\ref{a},\ref{b},\ref{d}) as in the bosonic case. Only Eq. (\ref{c})
has a slightly  different supersymmetric counterpart: 
 \be \delta_{\V}\{\M^2,Y^I\}=\{\M^2,\delta_{\V}Y^I\}= \{\ST(\V\M), \{\M^2,Y^I\}\}-i[\V,\{\M^2,Y^I\}].\label{cs}\ee
2. In order to speak about the   bosonic limit of (\ref{cor}), we must first   embed $\apo$ in $\sapo$ and then to consider the
action (\ref{cor}) evaluated at the configurations $\hat y^I\in\sapo$ which are the    images of bosonic configurations $y^I\in\apo$ upon  this embedding.  The embedding itself was
constructed in \cite{GKP} and it is completely defined by the images $\hat M_{ij}\in\sapo$  (denoted by "hats")  of the bosonic sphere  generators $M_{ij}\in\apo$:
\be \hat M_{ij}=(\M_{e}(1+\M_o^2))_{ij}, \quad i,j=1,2.\label{mat}\ee
Here $\M_e$ and $\M_o$   are  the   even  and the odd parts of the supermatrix $\M$ and it is perhaps useful to rewrite (\ref{mat}) in components: \be \hat M_{ij}=\M_{ij}(1+ \M_{13}  \bar \M_{13}).\ee
It is  easy to check that  (\ref{mat})
  is consistent with the sphere and supersphere defining relations (\ref{sph}) and (\ref{ssp}). Moreover, the embedding preserves the
  measure, i.e. it holds for every $f\in\apo$:
  \be \int \tro f=\int \std \hat f.\label{mea}\ee
  However, the Poisson structure is not completely preserved since it holds
  \be \{\hat f,\hat g\}_{\sus} =\widehat{\{f,g\}}_{S^{2}}(1+ \M_{13}  \bar\M_{13}), \quad f,g\in\apo.\label{bea}\ee
  The crux of the argument is now based
  on the following identities which can be verified straightforwardly from the definition (\ref{spb1}) of the supersphere Poisson structure:
  \be \{\M_e^2,\hat y\}=\{\M_o^2,\hat y\}=0,\quad \{\M_o\M_e,\hat y\}=\jp\M_o\{\M_e,\hat y\}, \quad \{\M_e\M_o,\hat y\}=\jp \{\M_e,\hat y\}\M_o, \label{ide}\ee
  where  $\hat y\in\sapo$ is the embedding of some  $y\in\apo$.
  We can now start to evaluate the action $S_{sGB}$ on the bosonic sigma model configuration $y^K\in\apo$ embedded in $\sapo$ as $\hat y^K$:
$$ S_{sGB}(\hat y^K)=-\ST\int \std (G_{IJ} +2\ri\M B_{IJ} ) \{\M^2,\hat y^I\}\{\M^2,\hat y^J\}=$$$$=-\ST\int \std (G_{IJ} +2\ri\M B_{IJ} ) \{\M_e\M_o+\M_o\M_e,\hat y^I\}\{\M_e\M_o+\M_o\M_e,\hat y^J\}=$$$$=-\js\ST\int \std (G_{IJ} +2\ri\M_e B_{IJ} ) (  \M_o\{\M_e,\hat y^I\}\{\M_e,\hat y^J\}\M_o+\{ \M_e,\hat y^I\}\M_o^2\{\M_e,\hat y^J\})=$$\be =-\jp\ST\int \std \M_o^2 (G_{IJ} + \ri\hat M B_{IJ} )\{ \hat M,\hat y^I\}\{\hat M,\hat y^J\}=\jp\Tr\int \tro (G_{IJ} +\ri MB_{IJ}) \{M,y^I\}\{M,y^J\}.\label{pa}\ee
In deriving (\ref{pa}) we have used (\ref{mea}),(\ref{bea}), the fact   that $\M_o\M\M_o=0$,   that $\M_o^2$ commutes with all matrices in the r.h.s. of (\ref{pa}) and also  the facts like  e.g. $\{\M_e\M_o,\hat y^I\}\{\M_e\M_o,\hat y^J\}$ vanishes being the  product of two odd upper-triangular matrices.  Morevover, the last equality in (\ref{pa}) is obtained by integrating over the odd generators $\M_{13},\bar \M_{13}$  which are present only
in the matrix $\M_o^2$  since at every other place, including the measure delta function $\delta(\M_{11}^2+  \M_{12}\bar \M_{12}+2\M_{13}\bar \M_{13}-1)$, they are killed by the nilpotency.

 \noindent 3. It remains to verify  the absence of a  quadratic  bosonic derivatives in the fermionic part of the action (\ref{cor}). For that we need not   enter two far into the jungle of component calculations.  We just
 consider the fermionic  part $Y^I_o$ of the superfield $Y^I$ and we can write it as
 \be Y^I_o(\M)=\Psi^I_-( \hat M)\M_{13}- \Psi_+^I( \hat M)\bar \M_{13},\ee
 where the matrix $\hat M$ was defined in (\ref{mat}) nad the reality of the superfield  $Y^I_o(\M)$ is ensured by requiring $\bar\Psi_+=\Psi_-$.
 Now we study the expression $\{\M^2,Y^I_o\}= \{\M^2,\Psi^I_-(\hat M)\M_{13}- \Psi^I_+(\hat M)\bar \M_{13}\}$ appearing in the action 
 (\ref{cor}).
We  find from Eq. (\ref{pb}) that the components
 of the even part $(\M^2)_e$ of the supermatrix $\M^2$ Poisson-commute with the components of $\M_e$.  This means that   only
 the Poisson bracket $\{(\M^2)_o, Y^I_o\}$ can contain the bosonic derivatives $\{\M_e,\Psi^I_{\pm}\}$. By using Eq. (\ref{ide}), we obtain  
 \be \{(\M^2)_o,\Psi^I_\pm(\hat M)\}= \{\M_o\M_e+\M_e\M_o,\Psi^I_\pm(\hat M)\}=\jp \M_o\{\M_e,\Psi^I_\pm\}-\jp\{\M_e,\Psi^I_\pm\}\M_o\ee
 hence
\be  \{(\M^2)_o, Y^I_o\}=  -\Psi^I_-(\hat M) \{(\M^2)_o,\M_{13}\}
+\Psi^I_+(\hat M)\{(\M^2)_o,\bar\M_{13}\}+ \M_{13}\bar \M_{13}V \ee
This means that the bosonic derivatives of fermions  $\{\M_e,\Psi^I_\pm\}$ appear only in the expression $V$ that multiplies $\M_{13}\bar \M_{13}$. The fact that $\M_{13}\bar \M_{13}$ squares to zero thus excludes a presence of the  quadratic  bosonic derivatives in the fermionic part of the action.

\subsection{Supersymmetric sigma model on $S^{2\vert 2}$ in components}

This paper would not be complet, if we did not present the component action derived from   the superfield action (\ref{cor}) via the ansatz
\be Y^I(\M)=\hat y^I+ \Psi^I_- \M_{13}- \Psi_+^I\bar \M_{13}+ F^I\M_{13}\bar \M_{13}.\ee
By eliminating the auxiliary fields $F^I$, we obtain
$$ S_{sGB}=\jp\int \tro \Tr\biggl[  (G_{IJ} +\ri MB_{IJ} ) \{M,y^I\}\{M,y^J\}+ 2G_{IJ}(\ri \{M,\Psi^J\}+\Psi^J)\bar\Psi^{I}\biggr]+$$
\be + \int\tro \biggl[\bar\Psi^{I}\ri \{M,y^K\}(\Gamma_{IKL}+\ri MH_{IKL})\Psi^L-\frac{1}{8} {\cal R}_{IJKL}(\bar\Psi^I\Psi^K-\bar\Psi^IM\Psi^K) (\bar\Psi^J\Psi^L-\bar\Psi^JM\Psi^L)\biggr],\label{comp}\ee
where
\be \Psi^I:= \left(\begin{matrix} \Psi^I_+\cr \Psi^I_-  \end{matrix}\right), \quad  \bar\Psi^I:= \left(\begin{matrix} \bar\Psi^I_+&\bar\Psi^I_-  \end{matrix}\right)=    \left(\begin{matrix} \Psi^I_-&-\Psi^I_+  \end{matrix}\right)\ee
and the notation $\{M,\Psi^J\}$ means at the same time the Poisson bracket and the matrix action on a column vector:
\be \{M,\Psi\}_\alpha:=\sum_{\beta}\{M_{\alpha\beta},\Psi_\beta\}.\ee
Moreover,  the quantities $\Gamma_{IKL}$, $H_{IKL}$ and ${\cal R}_{IJKL}$ are defined as
$$ \Gamma_{IKL}=\jp(\partial_{L}G_{KI}+\partial_{K}G_{IL}-\partial_{I}G_{KL}), \quad  H_{IKL}=\jp(\partial_{I}B_{KL}+\partial_{K}B_{LI}+\partial_{L}B_{IK}),
$$
\be {\cal R}_{IJKL}:=G_{IM}{\cal R}^{M}_{~JKL},\quad
  {\cal R}^{M}_{~JKL}:=\partial_KE^M_{~LJ}-\partial_LE^M_{~KJ}+  E^M_{~KN} E^N_{~LJ}-
E^M_{~LN} E^N_{~KJ},\ee
\be E^K_{~LJ}:=G^{KN}(\Gamma_{NLJ}+\ri H_{NLJ}).\ee
We recognize in the quantity $G^{KN} \Gamma_{NLJ}$ the standard Christoffel symbol corresponding to the metric $G_{IJ}$,
the totally antisymmetric tensor $H_{IJK}$  is nothing but the exterior derivative of the two-form $B_{IJ}$ and ${\cal R}_{IJKL}$
are the components of the modified Riemann curvature tensor corresponding to the connection $E^K_{LJ}$ containing the torsion
part $H^K_{LJ}$.

\noindent We notice that the component action (\ref{comp}) has again the elegant property that, apart from  the dynamical fields $y^I$ and $\Psi^I$,
it contains just the Poisson brackets and the moment map $M$. However, we have to admit that in this particular case we were not able to  preserve the elegance 
  in all intermediate calculations. Indeed, while everything else  in this paper was computed very directly and effortlessly thanks to our 
invariant Poisson language,  the formula 
(\ref{comp})   was  worked out  by a tedious component calculation.

 \section{Conclusions and outlook}
 Apart from our main result which is  the contruction of the action of the $UOSp(2\vert 1)$ supersymmetric sigma model with the Kalb-Ramond term, the present
 article also  offers a   conceptual simplification and  a technical streamlining of the results of the reference \cite{GKP}. In particular,
 all five generators $R_3,R_\pm,V_\pm$ of the Lie superalgebra $\uosp$ and all three supersymetric covariant derivatives $\Gamma,D_\pm$ appearing explicitely in majority of formulas of  \cite{GKP} are conveniently arranged as matrix elements of a single supermatrix $\M$ and its square $\M^2$.
 Moreover, all calculations of \cite{GKP} can be  rephrased in terms of the matrices $\M$ and $\M^2$ as a whole without  a necessity
 to manipulate the matrix elements themselves. As for the outlook, we expect that the construction of supersymmetric gauge theories on the supersphere presented
 in \cite{K} could be equally streamlined and rendered conceptually more transparent  by using the moment map supermatrix $\M$.

\end{document}